\documentclass{article}
\pdfoutput=1
\usepackage{graphicx,xspace,url}
\usepackage{booktabs,color}
\usepackage{listings}
\usepackage{trcover}

\usepackage{makeidx,graphicx,booktabs,xspace,url,listings,color,multicol,pifont}

\newcommand {\ff}{FastFlow\xspace}

\title{Accelerating sequential programs using FastFlow and self-offloading}
\trnum{TR-10-03}
\author{Marco Aldinucci\thanks{Computer Science Department, University
    of Torino, Italy. Email: adinuc@di.unito.it} \and Marco Danelutto
  \and   Peter Kilpatrick\thanks{Computer Science Department, Queen's
    University Belfast, UK.}  \and Massimiliano Meneghin\thanks{IBM Technology Campus, Dublin Software Lab, Ireland.} \and Massimo Torquati}
\date{February 12th, 2010}

\begin{document}
\maketitle
\begin{abstract}
FastFlow is a programming environment specifically targeting
cache-coherent shared-memory multi-cores. FastFlow is implemented as a
stack of C++ template libraries built on top of lock-free (fence-free)
synchronization mechanisms. In this paper we present a further
evolution of FastFlow enabling programmers to offload part of their workload
on a dynamically created software accelerator running on
unused CPUs. The offloaded function can be easily derived from pre-existing
sequential code. We emphasize in particular the effective trade-off between human productivity and
execution efficiency of the approach.
\paragraph{Keywords} Multi-core, parallel programming, streaming,
skeletons, accelerator, non-blocking, synchronization, lock-free, function offload.
\end{abstract}

\section{Introduction}
The entire hardware industry has been moving to multi-core, which
nowadays equips the large majority of computing platforms. The rapid shift
toward multi-core technology has many drivers that are likely to
sustain this trend for several years to come. In turn, software technology is also
responding to this pressure \cite{patterson:cacm:09}. Certainly, in the
long term, writing parallel programs that are efficient,
portable, and correct must be no more onerous than writing such programs for sequential
computers. To date, however, parallel programming has not
embraced much more than low-level communication and synchronization
libraries. In the hierarchy of abstractions, it is only slightly above
toggling absolute binary in the front panel of the machine. We
believe that, among many, one of the reasons for such failure is
the fact that programming multi-core is still perceived as a branch
of high-performance computing with the consequent excessive focus on
absolute performance measures. By definition, the raison d'{\^e}tre
for high-performance computing is high performance, but MIPS, FLOPS
and speedup need not be the only measure. Human productivity, total
cost and time to solution are equally, if not more, important
\cite{blog:acm:reed:2009}.
While a substantial methodological change will be required to allow
effective design of parallel applications from scratch, the shift to multi-core
is required to be graceful in the short term: existing applications should be ported to
multi-core systems with moderate effort (despite the fact that they could be redesigned
with larger effort and larger performance gain).

In this paper we present the \emph{FastFlow accelerator}, i.e. a software
accelerator based on \emph{FastFlow} lock-free programming
technology, and a methodology enabling programmers to seamlessly (and
semi-automatically) transform a broad class of existing C/C++ program
to parallel programs. The FastFlow software accelerator,
in contrast with classic hardware accelerators, allows execution of streams of
tasks on unused cores of the CPU(s).

The FastFlow accelerator is build on top of the \emph{FastFlow}
programming environment, which is a stack of C++ template libraries
that, conceptually,  progressively abstract the shared memory
parallelism at the level of cores up to the definition of useful
programming constructs and patterns (skeletons)
\cite{fastflow:web,ske:wikipedia,cole:manifesto:02}. Skeletons subsume a well-defined
parallel semantics, which is used to ensure the correctness of the
program when offloading tasks from a possibly sequential framework to
a parallel one. FastFlow is discussed in Sec. \ref{sec:fastflow}.

As we shall see in Sec. \ref{sec:fastflow:acc}, the FastFlow
accelerator ultimately consists in a specific usage of the FastFlow
framework. However, while FastFlow, in the general case, \emph{requires}
redesign of the application, the FastFlow accelerator suggests an easy and
rapid way to improve the performance of existing C++
applications. This is further reinforced by the relative popularity
(especially among non-specialists) of accelerator APIs, such as \emph{OpenCL},
\emph{CUDA},  IBM's Dynamic Application Virtualization \cite{ibm:offload:09}, and annotation
languages such as \emph{OpenMP}.
As we shall see in Sec. \ref{sec:perf} and Sec.~\ref{sec:exp},
one of the advantages of the FastFlow accelerator
with respect to these environments is the tiny overhead introduced by the
non-blocking lock-free synchronization mechanism which enables the
parallelization of very fine grain activities, and thus broadens the
applicability of the technique to legacy codes.
Finally, in Sec. \ref{sec:exp} we report on experiments with the proposed
technique using a couple of simple yet significant examples: the C++
Mandelbrot set application from Nokia TrollTech's QT
examples \cite{qt:web}, and a heavily hand-tuned C-based N-queens solver \cite{somers}.

\section{The \ff parallel programming environment}
\label{sec:fastflow}

\begin{figure}[t!]
\begin{center}
\includegraphics[width=\linewidth]{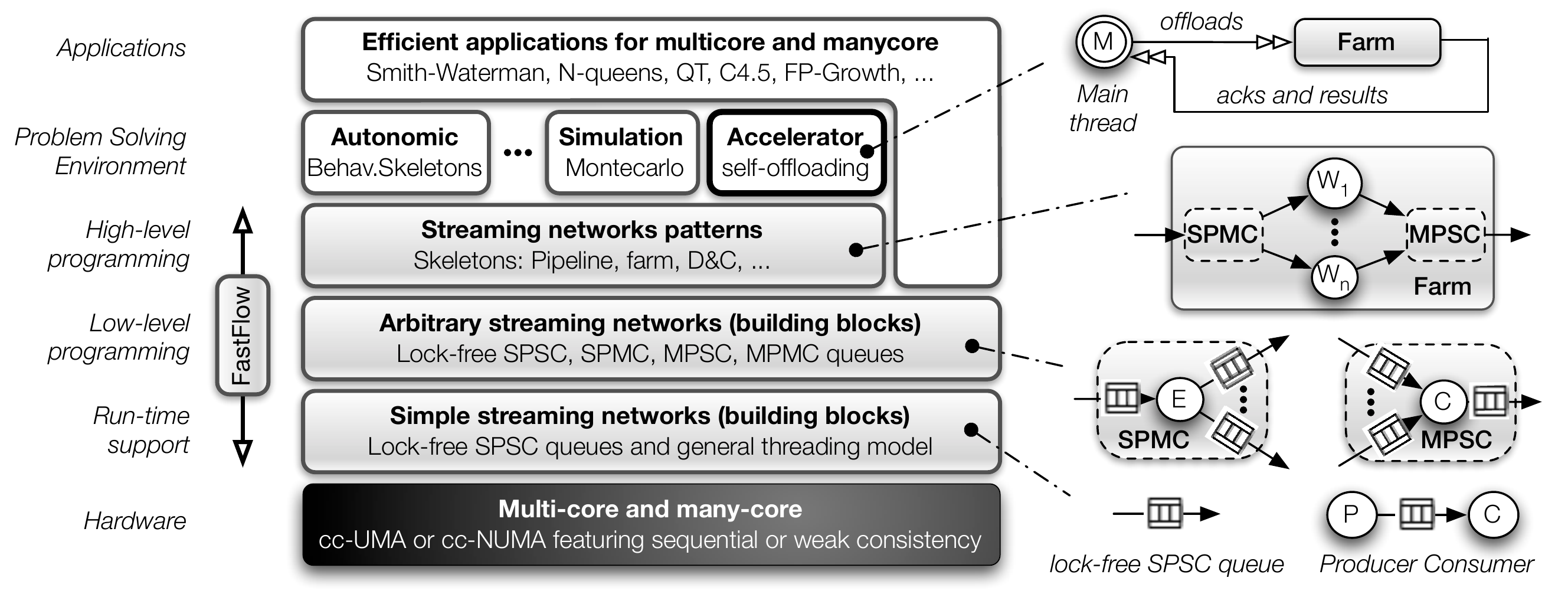}
\caption{FastFlow layered architecture with abstraction examples.\label{fig:ff:architecture}}
\end{center}
\end{figure}

As Fig.~\ref{fig:ff:architecture} shows, FastFlow is
conceptually designed as a stack of layers that
progressively abstract the shared memory parallelism at the level of
cores up to the definition of useful programming constructs and
patterns.  The abstraction process has two main goals:
1) to promote high-level parallel programming, and in particular skeletal
programming, i.e. pattern-based explicit parallel programming;
2) to promote efficient programming for multi-core.

\subsection{Hardware}
FastFlow is specifically designed for cache-coherent multiprocessors,
and in particular commodity homogenous multi-core (e.g. Intel core,
AMD K10, etc.).
It supports multiprocessors exploiting any memory consistency, including very
weak consistency models. FastFlow implementation is always
lock-free, and for several memory consistency models is also memory
fence-free  (e.g., sequential consistency, total store ordering, and
the x86 model). On other models (e.g., Itanium and Power4, 5, and 6),
a store fence before an enqueue is needed \cite{fastforward:ppopp:08}.

At the current development status, FastFlow does not include any
specific optimization targeting cc-NUMA platforms, although their
support is planned and under design. Also, it currently does not
automatically generate code running on GPUs (and other accelerators),
even if it admits and supports the linking of code running on hardware
accelerators. The full support of heterogenous platforms is currently
under evaluation.


\subsection{Run-time support}

\lstnewenvironment{Bench}[2]
{\lstset{ %
multicols=1,
language=C++,                
basicstyle=\scriptsize,       
numbers=none,                   
numberstyle=\tiny,      
stepnumber=1,                   
numbersep=2pt,                  
backgroundcolor=\color{white},  
showspaces=false,               
showstringspaces=false,         
showtabs=false,                 
tabsize=2,                      
captionpos=b,                   
breaklines=true,                
breakatwhitespace=false,        
escapeinside={\%*}{*)},          
mathescape=true,
columns=flexible,
morekeywords={main,add_workers,run_then_freeze, offload,offload_eos,load_result,wait_freezing},
#1,
label={code:#2},
showlines=true
}}{}

\begin{figure}
\centering

\begin{Bench}{}{}
bool push(void * const data) {
  if (!data) return false;
  if (buf[pwrite]==NULL) {
   //  WriteFence(); (e.g. for non x86 CPU)
    buf[pwrite] = data;
    pwrite+=(pwrite+1 >= size) ? (1-size): 1;
    return true;
  }
  return false;
}
bool  pop(void ** data) {
  if (!data || (buf[pread]==NULL))
    return false;
  *data = buf[pread];
  buf[pread]=NULL;
  pread += (pread+1 >= size)?(1-size): 1;
  return true;
}

\end{Bench}
\caption{Actual \ff SPCS \texttt{pop} and \texttt{push} C++ implementation.\label{fig:spsc}}
\end{figure}

Taking inspiration from Fastforward queues
\cite{fastforward:ppopp:08} and Lamport's wait-free protocols
\cite{Lamport}, the second tier provides mechanisms to define simple streaming
networks whose \emph{run-time support} is implemented through correct
and efficient lock-free Single-Producer-Single-Consumer (SPSC) queues.

The FastFlow run-time support layer realizes the two basic features:
\emph{parallelism exploitation}, i.e. the creation, destruction and life
cycle control of different flows of control sharing the
memory, and  \emph{asynchronous one-to-one communication channels}, supporting the
synchronization of different flows of control. They are implemented by
way of lock-free Single-Producer-Single-Consumer (SPSC) queue equipped
with non-blocking \texttt{push} and \texttt{pop} operations.

While the former point can be addressed using quite standard technology
(i.e. the wrapping of existing threading libraries, such as POSIX
threads), the second exhibits a number of performance pitfalls on
commodity shared-memory cache-coherent multiprocessors (as many
commodity multi-core are). In particular, traditional lock-free
implementations (such as Lamport's solution \cite{Lamport}) of SPSC queues are
correct under sequential consistency only, where none of the current
multi-cores implement sequential consistency. Also, some correct queue
implementations induce a very high invalidation rate -- and thus
reduced performance -- because they exhibit the sharing of locations
that are subject to alternative invalidations from communication
partners (e.g. head and tail of a circular buffers).
The implementation does not suffer from the ABA problem \cite{ABA:98}, and it remains
correct also in the case that only a reference instead of the full
message is communicated using the queues.  The FastFlow SPSC queue
implementation (shown in Fig.~\ref{fig:spsc}) is largely
inspired by Fastforward queues \cite{fastforward:ppopp:08}. As with
Fastforward queues,  the
\texttt{push} operation (issued by the   producer) always reads and writes \texttt{pwrite}
(i.e. tail pointer) only, and  the \texttt{push} (issued by the
consumer) always reads and writes \texttt{pread} (i.e. head pointer) only.
This approach substantially differs from the traditional one (e.g. in
Lamport's queues) where both the producer and the consumer access 
both the head and tail pointers causing the continuous invalidation
of cache lines holding head and tail pointers.



\subsection{Low-level programming}

One small, but significant, abstraction step is evident in the \emph{low-level
  programming} layer, which provides one-to-many, many-to-one, and
many-to-many synchronizations and data flows. In the FastFlow approach
these forms of communication are supported by SPMC
(Single-Producer-Multiple-Consumer), MPSC
(Multiple-Producer-Single-Consumer), MPMC
(Multiple-Producer-Multiple-Con\-sum\-er) queues, respectively. They
can be directly used as general asymmetric asynchronous channels among
threads.  Clearly, messages flowing through these channels may carry
memory pointers (that behave also as synchronization tokens), since we
are exploiting the underlying hardware
cache-coherent shared memory.
Abstractly, these queues realize a
general message passing API on top of a hardware shared memory
layer.

SPMC,  MPSC, and  MPMC queues can be realized in several different
ways, for example using locks, or  in a lock-free fashion in order to avoid lock
overhead (which is a non-negligible overhead
in multi-core architectures). However,
these queues could not be \emph{directly} programmed in a lock-free
fashion without using at least
one atomic operation, which is  typically used to enforce the correct
serialization of updates from either many producers or many consumers at the
same end of the queue.
These operations, however,  induce a memory fence, thus a
cache invalidation/update\footnote{Notice that building a lock also
  requires an atomic operation unless working under sequential
  consistency for which a number of algorithms that do not require
  atomic operations exist, e.g. Lamport's
  Bakery algorithm \cite{lamport:bakery}.}, which  can seriously
impair the performance of parallel
programs exhibiting frequent synchronizations (e.g. for fine-grain
parallelism).

With FastFlow we advocate a different approach to the
implementation of these queues, which require neither locks nor atomic operations.
SPMC, MPSC, and MPMC queues are
realized by using only SPSC queues and an arbiter thread, which enforce the correct
serialization of producers and consumers. As shown in
Fig.~\ref{fig:ff:architecture}, this arbiter thread is called \emph{Emitter}
(E) when it  is used to dispatch data from one channel to many channels, \emph{Collector} (C) when
it is used to gather data from many channels and push the messages into one
channel, and \emph{Collector-Emitter} (CE) when it behaves both as
Collector and Emitter  (a.k.a. Master-workers pattern).

Notice that, at this level, FastFlow does not make any decision
about thread scheduling and their mapping onto the core; the
programmer should be fully aware of all programming aspects and their
potential performance drawback, such as load-balancing and memory
alignment and hot-spots.

\subsection{High-level programming}
The next layer up, i.e. \emph{high-level programming},  provides a
programming framework based on parallelism exploitation patterns
(\emph{skeletons} \cite{ske:wikipedia}). They are usually categorized
in three main classes: Task, Data, and Stream Parallelism.
\ff specifically focuses on Stream Parallelism, and in particular provides:
\emph{farm}, \emph{farm-with-feedback} (i.e. Divide\&Conquer), \emph{pipeline},
and their arbitrary nesting and composition.
The set of skeletons provided by FastFlow could be further extended by building new
C++ templates.

Stream Parallelism can be used when there exists a partial or total
order in a computation. By processing data elements in order, local
state may be maintained in each filter.
The set of skeletons provided by FastFlow could be further extended by building new
C++ templates on top  of the Fastflow low-level programming layer.

Task Parallelism is explicit in the algorithm and consists of running
the same or different code on different executors (cores, processors,
machines, etc.). Different flows-of-control (threads, processes, etc.)
may communicate with one another as they work.
Communication usually takes place to pass data from one thread to the next as part of the same data-flow graph.

Data Parallelism is a method for parallelizing a single task by
processing independent data elements of this task in parallel. The
flexibility of the technique relies upon stateless processing routines
implying that the data elements must be fully independent. Data
Parallelism also supports Loop-level Parallelism where successive
iterations of a loop working on independent or read-only data are
parallelized in different flows-of-control and concurrently
executed.


While many of the  programming frameworks mentioned in
Sec.~\ref{sec:relwork} offer Data and Task Parallel skeletons, only
few of them offer Stream Parallel skeletons (such as TBB's
\emph{pipeline}). None of them  offers the \emph{farm} skeleton, which
exploits  functional replication of a set of \emph{workers} and  abstracts out the parallel
filtering of successive \emph{independent} items of the stream under
the control of a scheduler, as a first-class concept.

We refer to \cite{fastflow:web} for implementation details
and to \cite{fastflow:parco:09,fastflow:pdp:10} for a performance
comparison against POSIX locks, Cilk, OpenMP, and TBB.
\ff is available at
\url{http://sourceforge.net/projects/mc-fastflow/} under GPL.

\section{Self-offloading on the \ff accelerator}
\label{sec:fastflow:acc}

\begin{figure}
\begin{center}
\includegraphics[width=\linewidth]{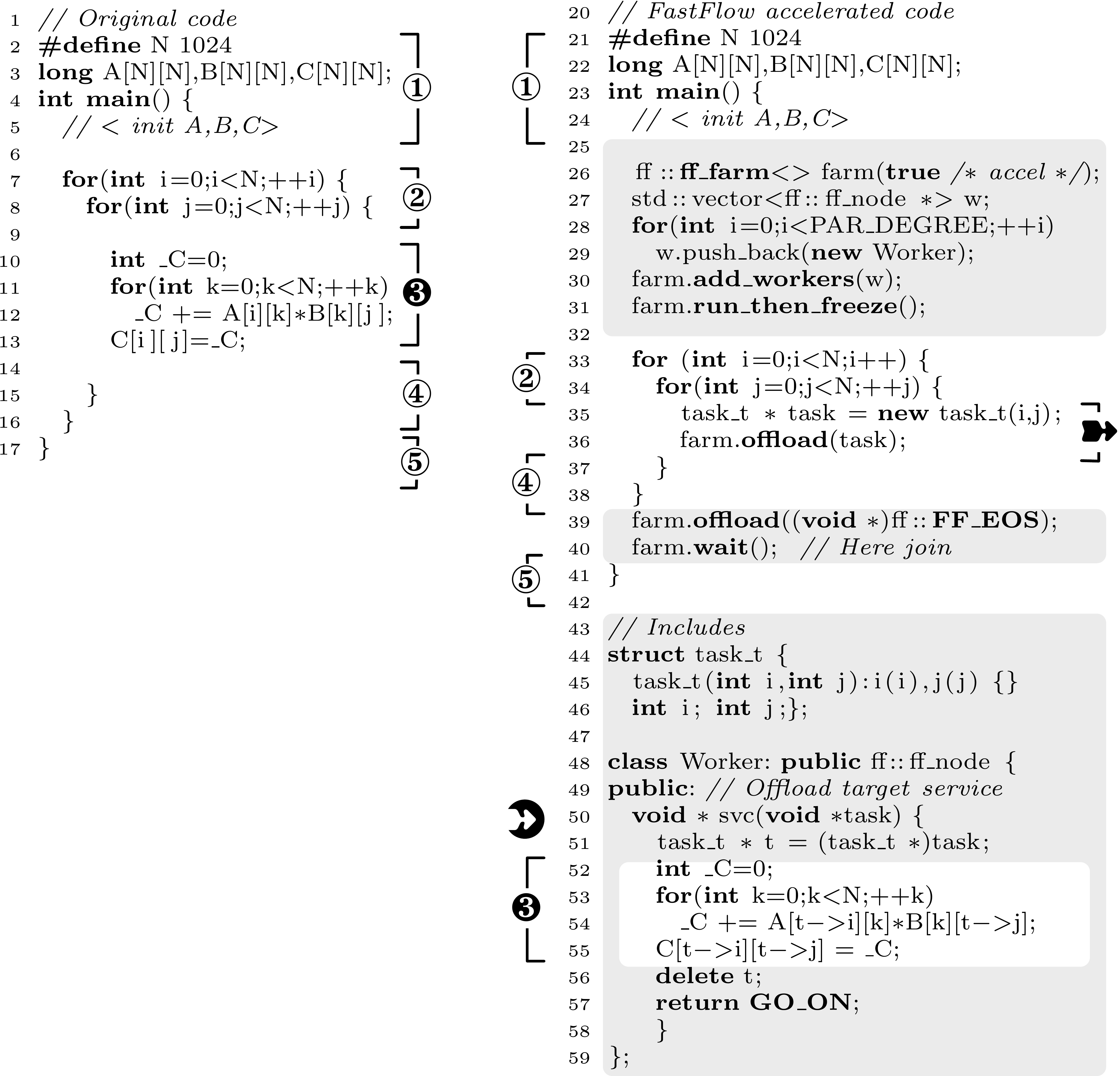}
\end{center}
\vspace{-36ex}
\begin{minipage}{34ex}
\scriptsize

Regions marked with white circled figures
  \ding{172},\ding{173},\ding{174},\ding{176} are
  copy-pasted.

The region marked with black circled figure (\ding{184}) has been
  selected to be accelerated with a farm. It is copied with
  renaming of  variables that are concurrently changed, e.g. automatic
  variables in a loop. A stream of task\_t variables is used to keep
  all different values of these variables.

Grey boxes create and run the accelerator; they are pre-determined
according to the accelerator type.

The code marked with \ding{253} executes the offloading onto the
  accelerator; the target  of  the offloading is the svc method {\small
    \ding{108}{\tiny$\!\!\!\!\!\!\!\!$}{\color{white}\ding{253}}} of the
  Worker class.
\end{minipage}
\vspace{9ex}
\caption{Derivation of FastFlow accelerated code from a simple
  sequential C++ application (matrix multiplication).}
\label{code:bench}
\end{figure}

A \emph{FastFlow accelerator} is a software device wrapping a
high-level \ff program, i.e. a skeleton or a composition of
skeletons, and providing the application programmer with a functional
\emph{self-offloading} feature, since the offload happens on the same hardware
device, i.e. CPU cores.
The primary aim of self-offloading is to provide the
programmer with an easy and semi-automatic path to introducing parallelism into a
C/C++ sequential code by moving or copying parts of the original code
into the body of C++ methods, which will be executed in parallel according
to a \ff skeleton (or skeleton composition). This
requires limited programming effort and it may speed up  the
original code by exploiting unused cores.

A FastFlow accelerator provides the programmer with one (untyped) streaming input
channel and one (untyped) streaming output channel that can be dynamically \emph{created} (and
\emph{destroyed}) from a C++ code (either sequential or multi-threaded) as
a C++ object (Fig.~\ref{code:bench} lines 26--30). Thanks to the underlying shared memory architecture, messages
flowing into these channels may carry both values and pointers to data structures.

An accelerator, which is a collection of threads,  has a global lifecycle with
two stable states: \emph{running} and
\emph{frozen}, plus several transient states. The running state happens
when all threads are logically able to run (i.e. they are ready or
running at the O.S. level).  The frozen state happens when all threads
are suspended (at the O.S. level). Transitions from these two
states involve calls to the underlying threading library (and to the O.S.).

Once created, an accelerator can be run (line 31), making it capable of
accepting tasks on the input channel. When running, the threads
belonging to an accelerator might fall into an \emph{active waiting}
  state. These state
transitions exhibit a very low overhead and do not involve the O.S.
Threads not belonging to the accelerator
could \emph{wait} for an accelerator, i.e. suspend until the accelerator completes its input tasks (receives the
\emph{End-of-Stream}, unique is propagated in transient states of the
lifecycle to all threads)  and then put it in the frozen state.
At creation time, the accelerator is configured and its threads are bound into
one or more  cores. Since the FastFlow run-time is implemented via non-blocking threads,
they will, if not frozen, fully load the cores in which they are
placed, no matter whether they are actually processing something or not.
Because of this, the accelerator is usually configured to use ``spare'' cores
(although over-provisioning could be forced).  If necessary, output tasks
could be popped from the accelerator output channel.

\subsection{Accelerating standard C++ codes: how to}
\label{sec:selfoffloading}

A \ff accelerator is defined by a  FastFlow skeletal
composition augmented with an input stream and an output stream that
can be, respectively, pushed and popped from outside the accelerator.
Both the functional and extra-functional behaviour  of the accelerator is
fully determined by the chosen skeletal composition. For example, the
\emph{farm} skeleton provides the parallel execution of the same  code
(within a \emph{worker} object) on independent items of the input
stream. The \emph{pipeline} skeleton provides the parallel execution
of filters (or stages) exhibiting a direct data dependency. More
complex behaviours can be defined by creating compositions of skeletons
\cite{beske:ipdps:09,lithium:sem:CLSS};
whose behaviour could be
described using (acyclic or cyclic) data flow graphs.  As we will see, clear knowledge of
accelerator behaviour makes it possible to correctly parallelize segments
of code.

The use of a farm accelerator is exemplified in
Fig.~\ref{code:bench}. The code in the left column of the
figure  (lines 1--17)  shows a sequential program including three loops: a simple matrix
multiplication. Its accelerated version is shown on the right column (lines
20--59). The accelerated version can be semi-automatically derived from the
sequential by copy-pasting pieces of code into placeholders on a code
template (parts in white background in the left column): for example, code marked with
\ding{172},\ding{173},\ding{175}, and \ding{176} are copied from left
to right. The code that has been selected for the offloading, in this case the body of a
loop marked with \ding{184}, is copied into the worker body after a
suitable \emph{renaming} of variables.

Because it is composed of threads, the accelerator shares the memory
with its caller (and other threads of the same process). As is well-known,
transforming a sequential program into a parallel one requires
regulation of possibly concurrent memory accesses. In low-level
programming models this is usually done by using critical sections
and monitors under the responsibility of the programmer. FastFlow does
not prohibit these mechanisms, but promotes a methodology to avoid
them. In very general terms, the sequential code statement can be correctly accelerated with \ff
only mechanisms if the offloaded code and the offloading code
(e.g. main thread) instances do not break any data dependency,
according to Bernstein's conditions.  \ff helps the programmer in
enforcing these conditions in two ways: \emph{skeletons} and \emph{streams}.

The \emph{skeletal} structure of the accelerator induces a well-defined  partial ordering
among offloaded parts of code. For example, no-order for farm,  a
chain of dependencies for pipeline, a directed acyclic graph for
farm-pipeline nesting/composition, and a graph for a
farm-with-feedback.
The synchronization among threads is enforced by \emph{streams} along the paths of the particular skeleton composition, as in a data-flow
graph. True dependencies (read-after-write) are admissible only along
these paths. Streams can carry values or pointers, which act as
synchronization tokens for indirect access to the shared memory.


Pragmatically, streams couple quite well with the needs of sequential
code parallelisation. In fact the creation of a stream to be offloaded
on the accelerator can be effectively used to resolve anti-dependency
(write-after-read) on variables since the stream can carry a copy of
the values. For example, this happens when an iteration variable of
an accelerated loop is updated after the (asynchronous) offload. This
case naturally generalizes to all variables exhibiting a
larger scope with respect to the accelerated code. The same argument can
be used for output dependency (write-after-write). \ff accelerator
templates accommodate all variables of this kind in one or more
structs or C++ classes (e.g. \texttt{task\_t}, lines 44--46) representing
the input, and, if present, the output stream data type. All other data accesses
can be resolved by just relying on the underlying shared memory
(e.g. read-only, as A  at line 54, and single assignment as C at
line 55). The general methodology to accelerate
existing C++ codes using the \ff accelerator is described in Table~\ref{tab:methodology}.
\begin{table}[t]
\textsf{\small
\rule{\linewidth}{\heavyrulewidth}
\vspace{-4ex}
\begin{enumerate}
\item Choose a part of the
code to be accelerated (e.g. a heavy kernel), understand the data dependencies,
e.g. loop with independent iterations, data
dependencies between functions or basic blocks, or more complex
dependencies.
\item Choose a skeletal composition that models the required parallel
  execution schema.
\item Copy
and paste the chosen code into the accelerator parts according to the
skeleton template, e.g. in the farm worker,
emitter (data scheduling), collector (data gathering and
reduction).
\item Update accelerated code to access the memory via either
stream values or pointers, if necessary.
\item Fill the skeleton
template with accelerator creation and management code.
\item Substitute accelerated code with offloading calls.
\end{enumerate}
}
\vspace{-2.5ex}
\rule{\linewidth}{\heavyrulewidth}
\vspace{1ex}
\caption{Self-offloading methodology.}
\label{tab:methodology}
\end{table}

It is worth noticing that the \ff acceleration methodology may not be fully
automated. It has been conceived to ease the task of parallelisation by
providing the programmer with a methodology that helps in dealing with
several common cases.
However, many tasks require the programmer to make decisions, e.g. the selection of
the code to be accelerated. In the example code in Fig.~\ref{code:bench}
there are several choices with different computation granularity: offload only the index $i$ or the indexes $i$ and $j$, or all three indexes.
Also, the correctness of the final code depends on the
programmer: they should ensure that the accelerated code is thread
safe, streams have a suitable type and their pointers are correctly
cast, memory accesses are properly renamed, etc.
\ff,  like C/C++ itself, gives to the programmer much
flexibility that should be used with great care.

\subsection{Effectiveness and performance}
\label{sec:perf}
The \ff accelerator aims to provide good speedup with moderate
effort. Applications accelerated with \ff, in contrast with
fully-fledged \ff applications, are not \emph{fully} parallel. As with the
other accelerators, Amdahl's law applies. Thus, the maximum speedup
of an accelerated application depends primarily on which parts of the
code have been offloaded, and on what fraction of the overall execution
time is spent in that code. Equally important for performance is the
quality of the parallel code running on the accelerator in terms of
computation vs communication size, load balancing, memory
alignment, data locality and avoidance of false-sharing. For these problems
\ff provides the programmer with specific tools to tune the performance:
a parallel memory allocator, mechanisms to control
task scheduling, and a mechanism to trace the execution of the workers' threads.
A description of these tools goes beyond
this paper: we refer to the \ff documentation for further details \cite{fastflow:web}.

A significant advantage of the \ff accelerator with respect to other tools
is the low latency of the run-time and the high flexibility of the framework.
This, in turn, widens the parallelization possibilities to a broader class of
applications, and especially those programs performing frequent synchronizations
(e.g. fine-grain parallelism).

\section{Experiments}
\label{sec:exp}

In this section we show the performance of the FastFlow Accelerator
using two well-known applications: a Mandelbrot set explorer and an N-queens solver.
In both cases the code is third party and has been designed as
sequential code; then it has been made parallel with the
\ff farm accelerator.
All experiments reported in the following sections have been executed
on two platforms:
\begin{description}
\item[Andromeda] Intel  workstation with 2 quad-core Xeon E5520
Nehalem (16 HyperThreads) @2.26GHz with 8MB L3 cache
and 24 GBytes of main memory.
\item[Ottavinareale] Intel workstation with 2 quad-core Xeon E5420
Harpertown @2.5GHz with 6MB L2 cache and 8 GBytes of
main memory.
\end{description}
With the exception of very long runs, all
presented experimental results are taken as an average of 5 runs exhibiting
very low variance. All tested codes are available at the \ff website \cite{fastflow:web}.

\subsection{Interactive Mandelbrot set application}

\begin{figure}
\begin{center}
\rule{\linewidth}{\heavyrulewidth}
\textsf{\small Tests on Andromeda: 8-core 16-hyperthreads Intel platform}\\
\parbox{0.49\linewidth}{\hspace{2ex}
\textsf{\tiny
\begin{tabular}[b]{r@{\hspace{2px}}r@{\hspace{2px}}r@{\hspace{2px}}r@{\hspace{2px}}r}
\toprule
\multicolumn{5}{c}{Original QT-Mandelbrot execution time breakdown (mS)}\\
\midrule
Pass & Base & Wreath &2-Helix & Broccoli\\
\cmidrule{2-5}
&\includegraphics[scale=.10]{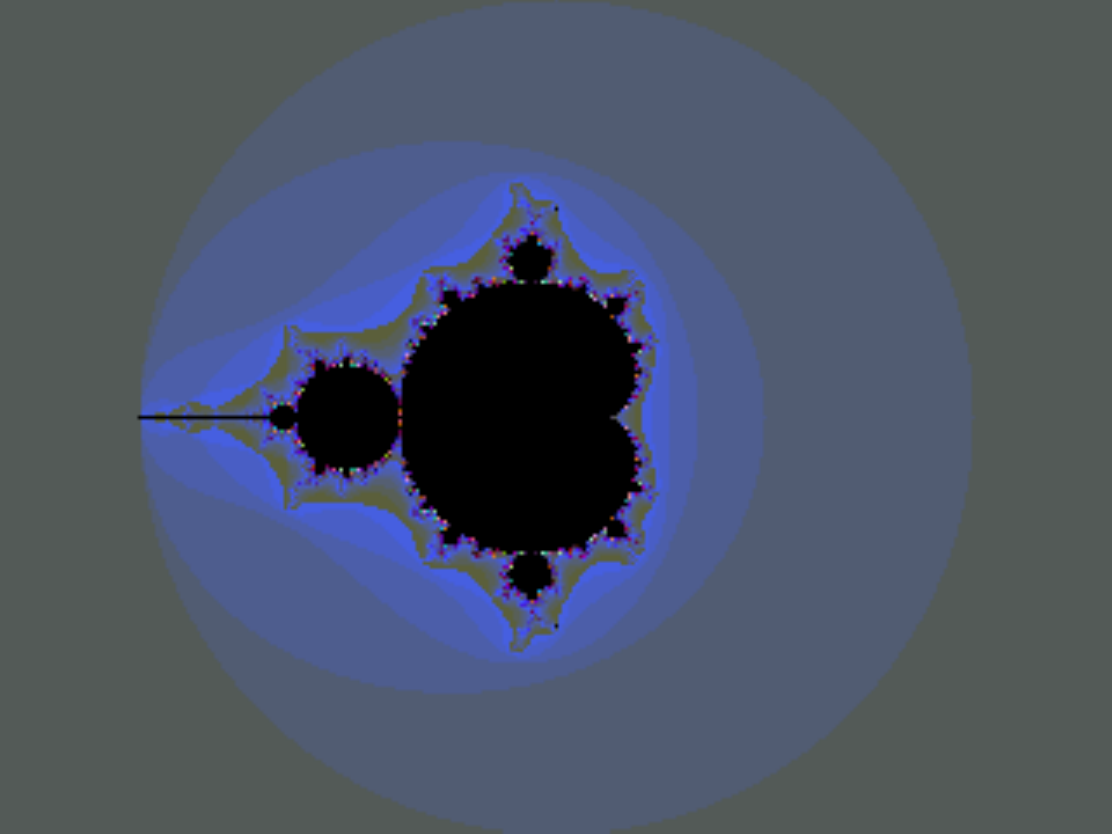}
&\includegraphics[scale=.10]{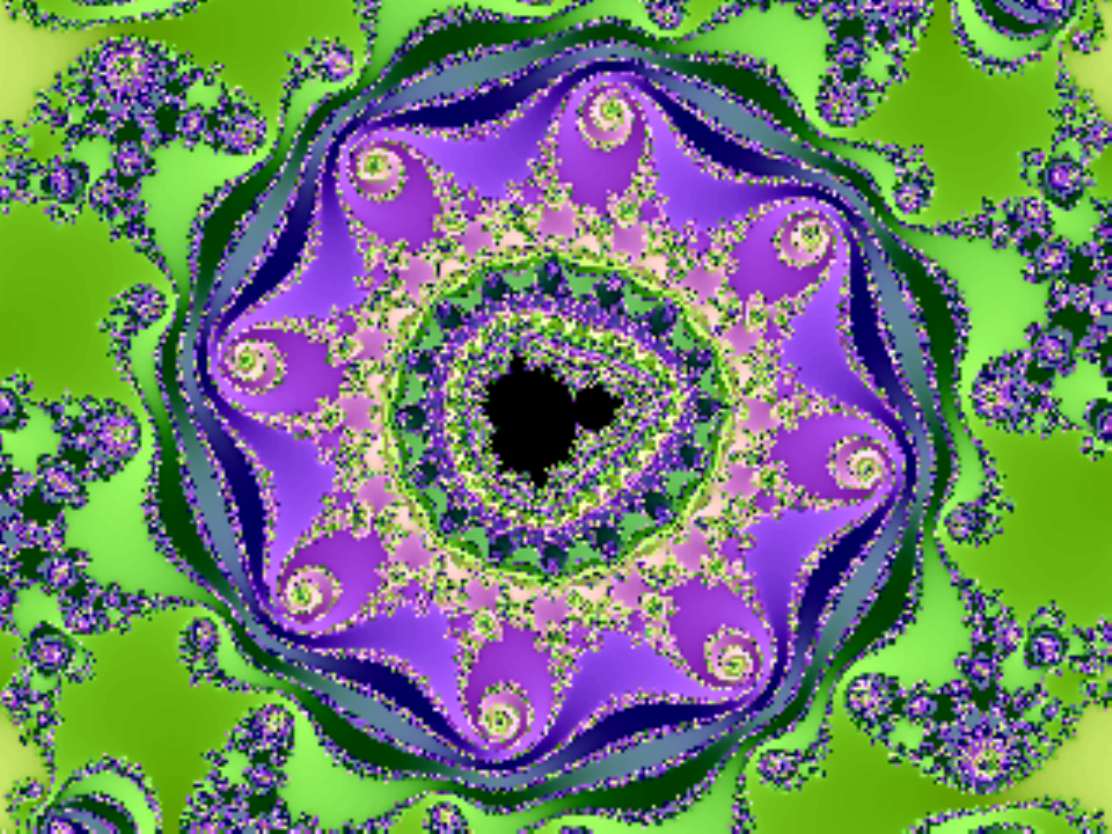}
&\includegraphics[scale=.10]{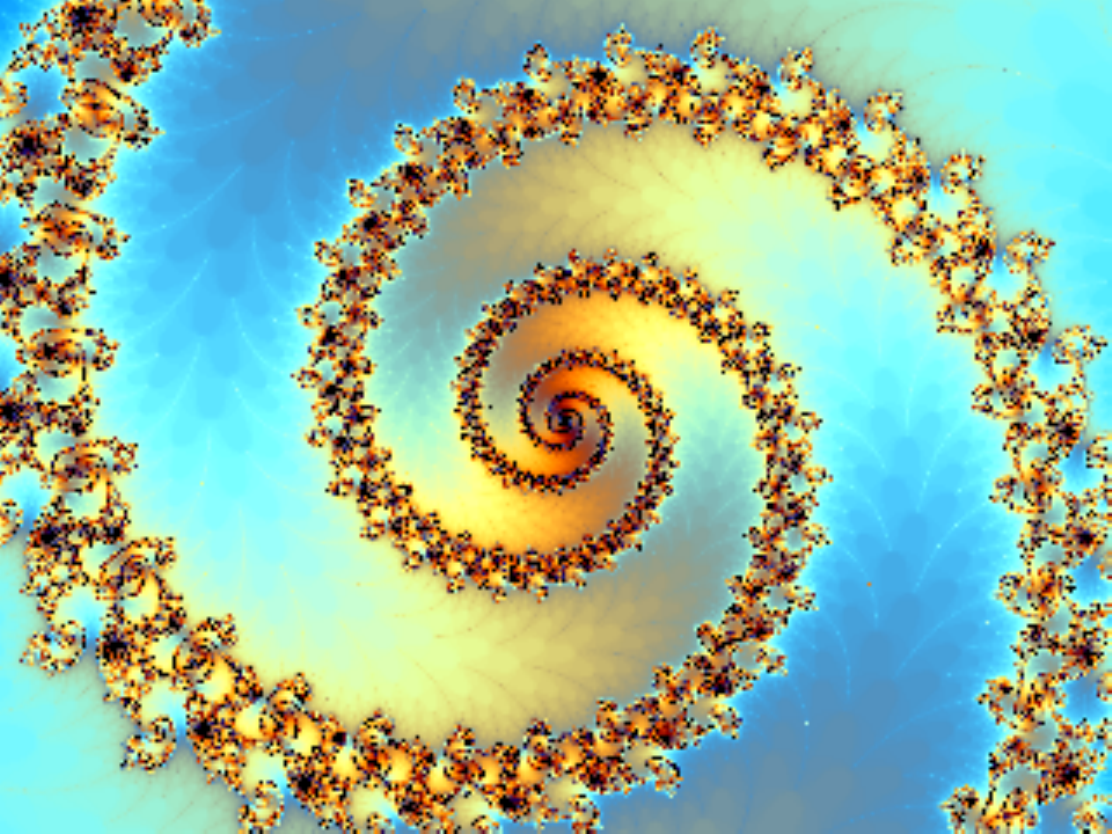}
&\includegraphics[scale=.10]{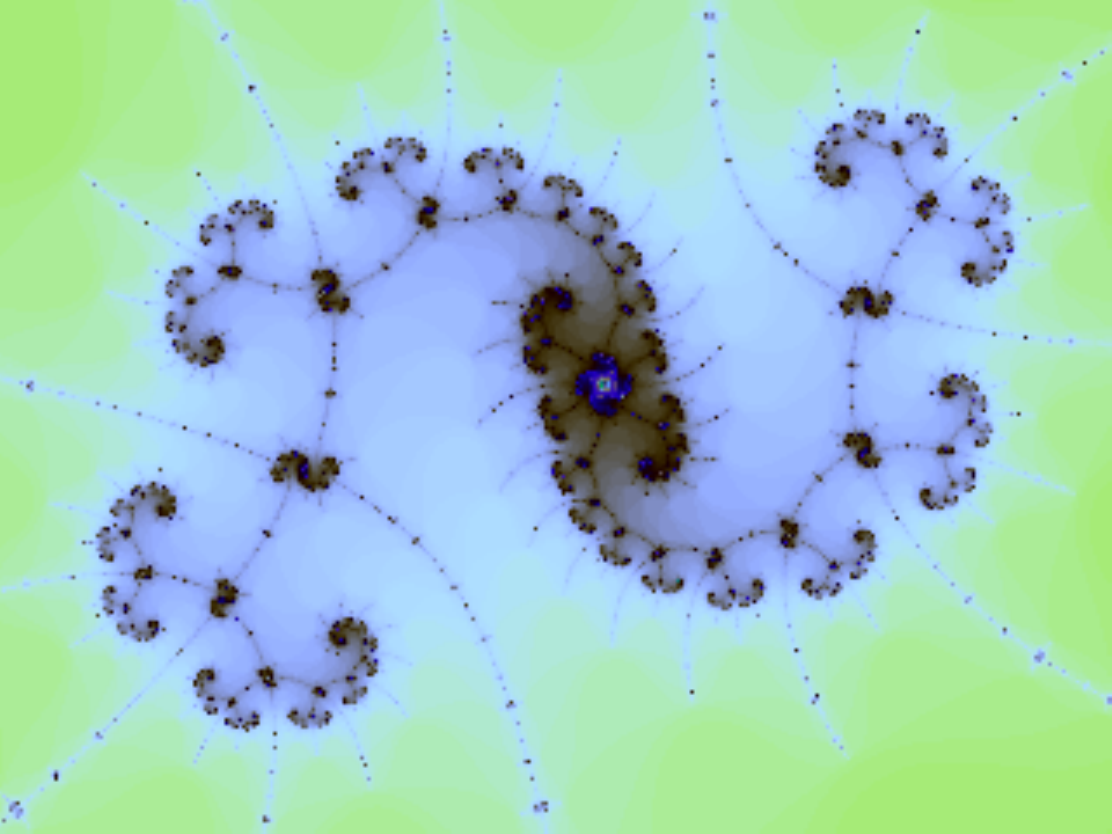}\\
\cmidrule{2-5}
1& 79&      140&  154&  128 \\
2& 98&	    382&  435&  120 \\
3& 696&	    1364& 651&  120 \\
4& 2682&    1526& 663&  121 \\
5& 10430&   1570& 670&  120 \\
6& 42203&   1722& 672&  121 \\
7& 168532&  2312& 682&  120 \\
8& 673203&  4673& 721&  121\\
\bottomrule
\end{tabular}
}}
\parbox{0.5\linewidth}{
\includegraphics[width=\linewidth]{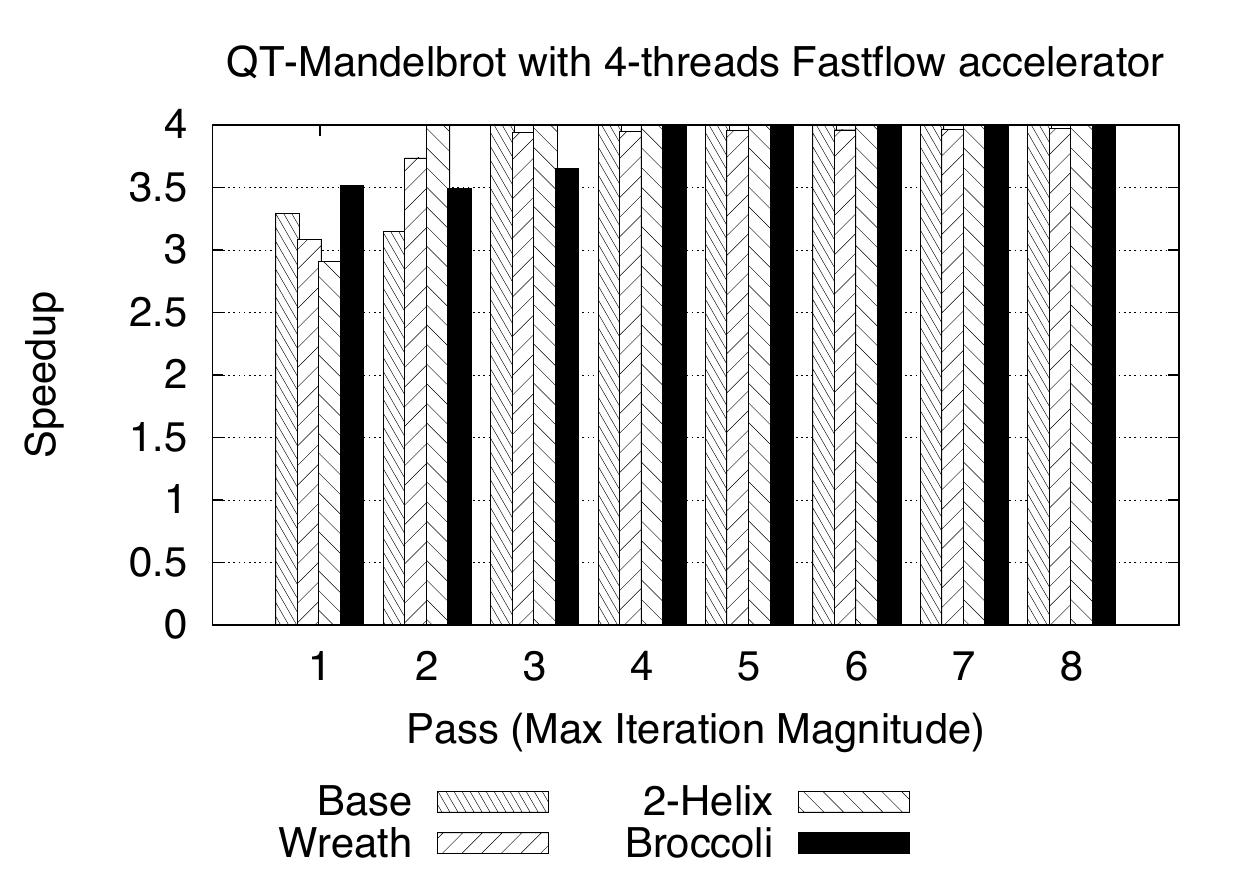}
}\\
\includegraphics[width=0.49\linewidth]{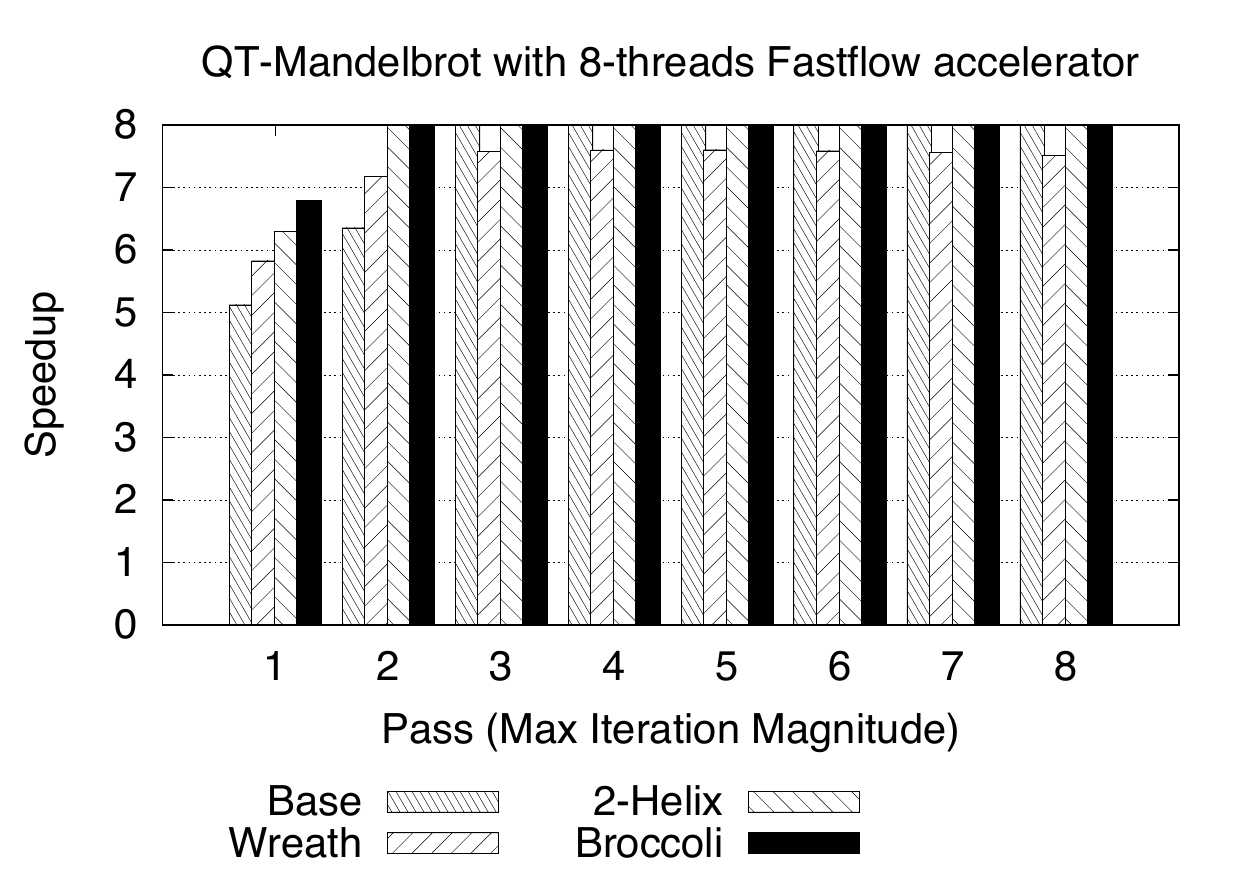}
\includegraphics[width=0.49\linewidth]{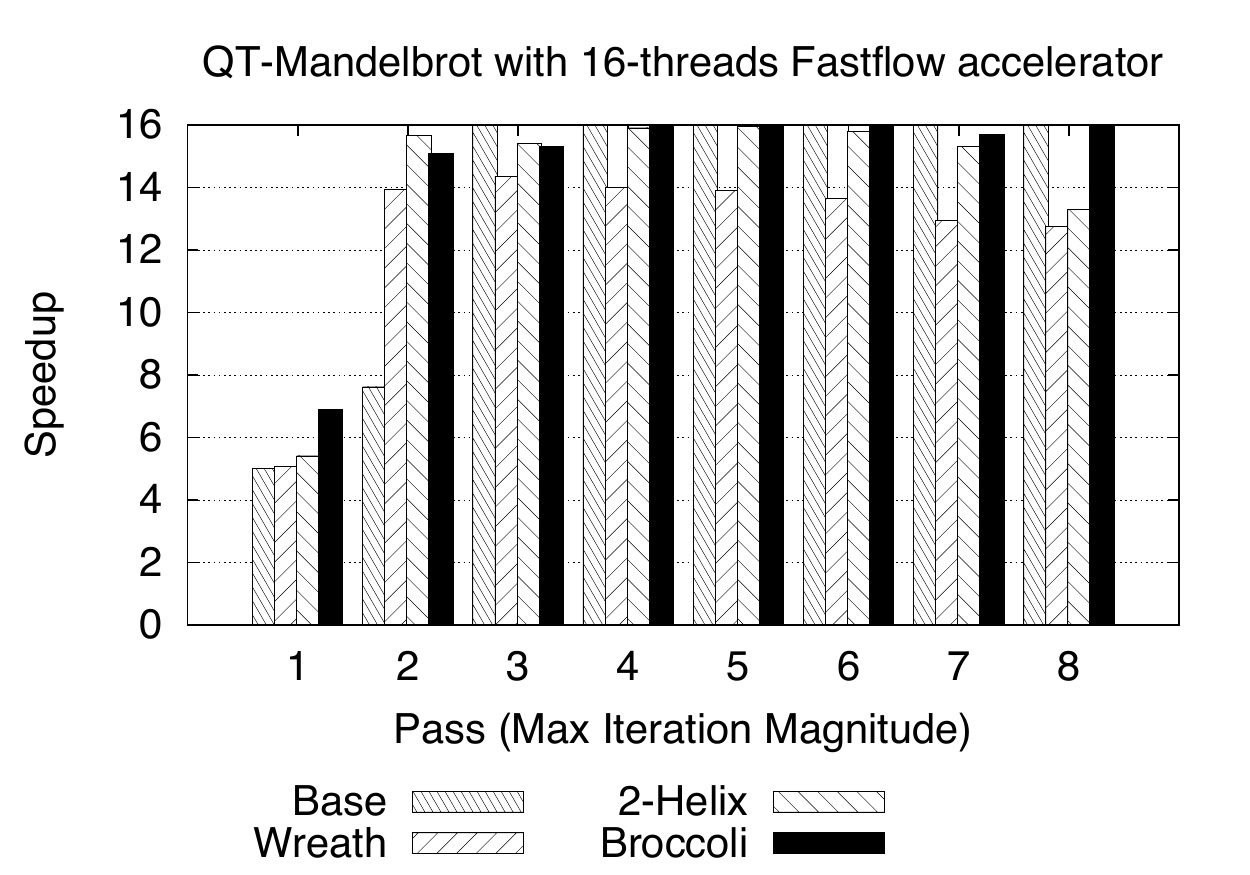}


\rule{\linewidth}{\heavyrulewidth}
\textsf{\small Tests on Ottavinareale: 8-core Intel platform}\\
\parbox{0.49\linewidth}{\hspace{2ex}
\textsf{\tiny
\begin{tabular}[b]{r@{\hspace{2px}}r@{\hspace{2px}}r@{\hspace{2px}}r@{\hspace{2px}}r}
\toprule
\multicolumn{5}{c}{Original QT-Mandelbrot execution time breakdown (mS)}\\
\midrule
Pass & Base & Wreath &2-Helix & Broccoli\\
\cmidrule{2-5}
&\includegraphics[scale=.10]{standard}
&\includegraphics[scale=.10]{wreath}
&\includegraphics[scale=.10]{2-helix}
&\includegraphics[scale=.10]{broccoli}\\
\cmidrule{2-5}
1&      64&      133&   133&    106\\
2&      177&    398&    391&    107\\
3&      617&    1422&   586&    107\\
4&      2370&   1594&   594&    107\\
5&      9379&   1638&   595&    107\\
6&      37437&  1795&   597&    107\\
7&      150025& 2413&   606&    107\\
\bottomrule
\end{tabular}
}}
\parbox{0.5\linewidth}{
\includegraphics[width=\linewidth]{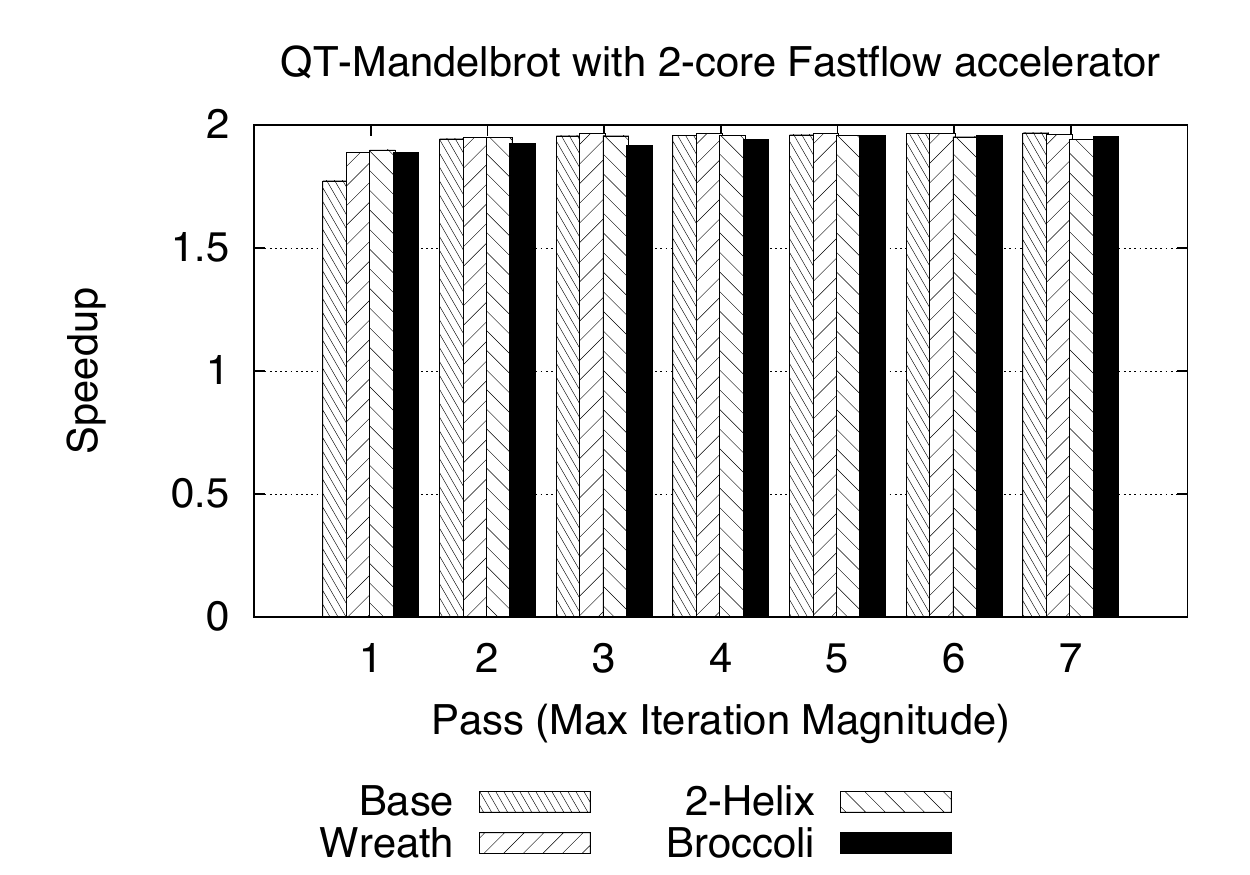}
}\\
\includegraphics[width=0.49\linewidth]{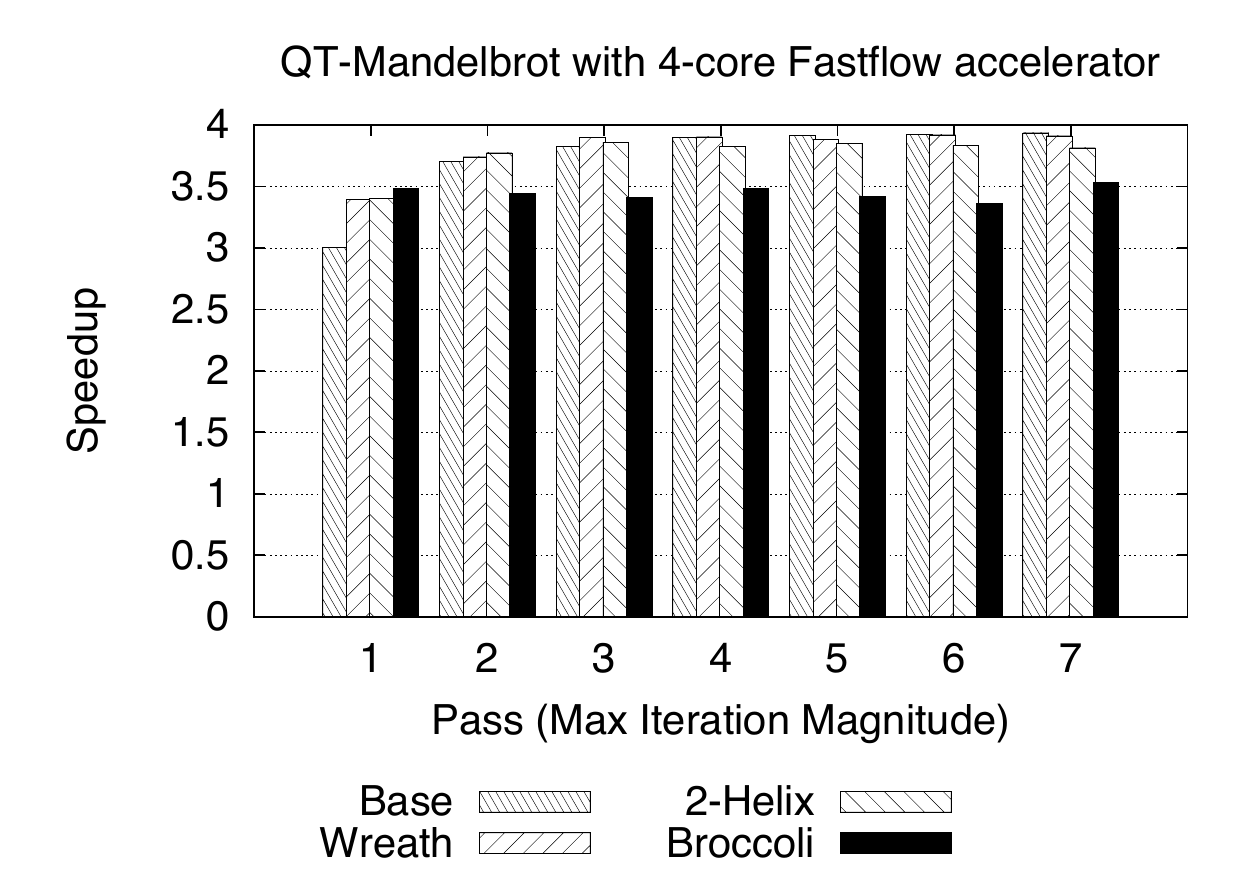}
\includegraphics[width=0.49\linewidth]{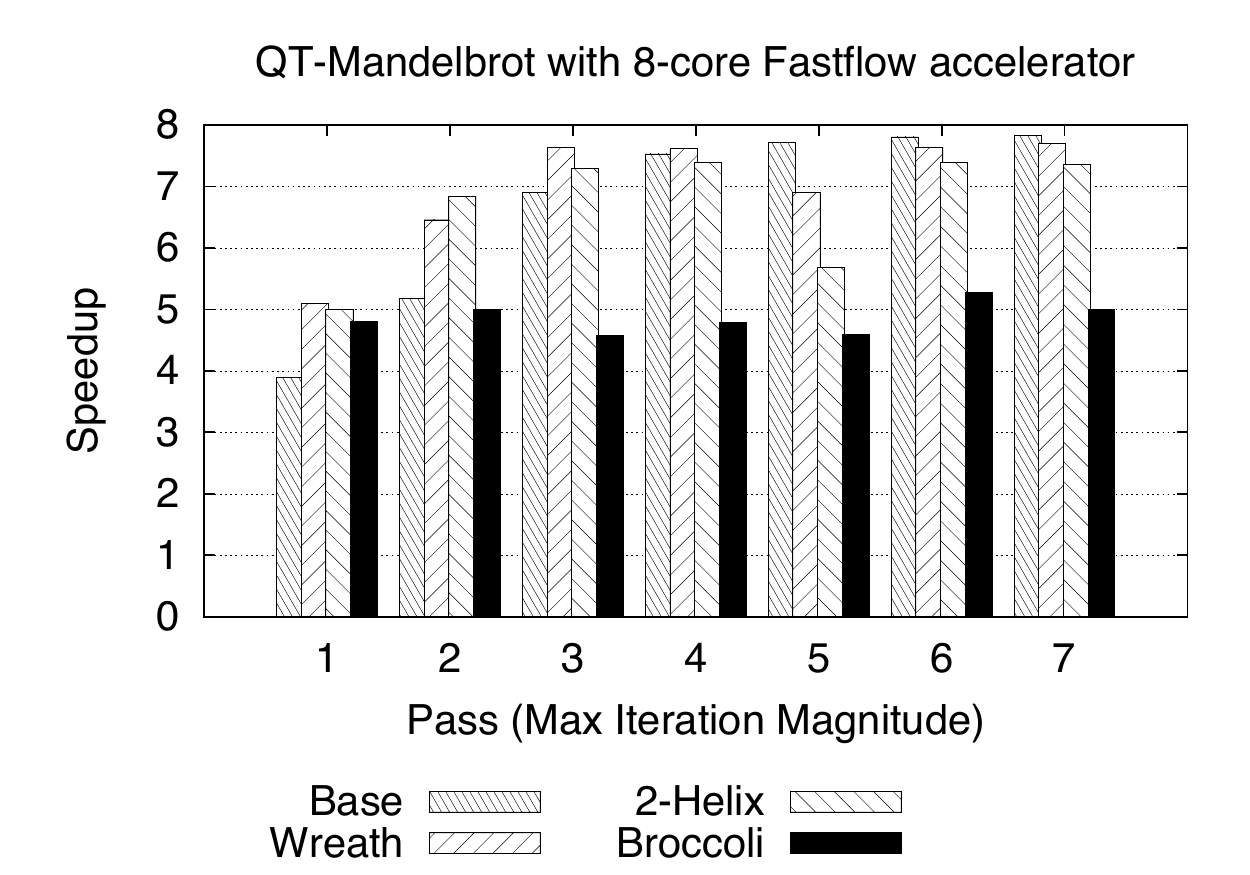}
\end{center}
\vspace{-2ex}
\caption{Original QT-Mandelbrot execution time with progressive
  precision (passes) in 4 different regions of the Mandelbrot set
  and the speedup obtained with
\ff on two multi-core platforms (Andromeda and Ottavinareale).\label{fig:mandelbrot}}
\end{figure}

The ``QT Mandelbrot''  is an interactive application that computes the
Mandelbrot set  \cite{qt:web}.
It is part of the Trolltech QT examples and it consists of two
classes: \texttt{RenderThread.cpp}, i.e. a \texttt{QThread} subclass that renders the
Mandelbrot set, and \texttt{MandelbrotWidget.cpp}, a \texttt{QWidget}
subclass that shows the Mandelbrot set on screen and lets the user
zoom and scroll.
The application is multi-threaded (the two classes are run as
QT threads) but threads are not used to speed
the computation up since the whole computation is done within a single
thread; rather they are used to decouple two different activities and to enhance
responsiveness. During the time when the worker thread is
recomputing the fractal to reflect the new zoom factor position, the
main thread  scales the previously rendered pixmap to provide
immediate feedback. This use of threads is quite common in
real life applications,
where the user interface must remain responsive while
some heavy operation is taking place.

Although it is a well-understood problem, the fully-fledged parallelization of
the whole application is not trivial. The two threads synchronise with
each other via QT events; to guarantee responsiveness the
\texttt{MandelbrotWidget} thread may start, restart, and abort the
activity of \texttt{RenderThread}. This kind of behavior, as well as the
integration of QT threads with other threading libraries, makes
porting to frameworks such as TBB and OpenMP non-trivial.
The \ff accelerated version makes parallel the
\texttt{RenderThread} by using a farm accelerator on the outer
loop that traverses the pixmap of the Mandelbrot set. The farm
accelerator is created once, then run and frozen each
time a compute and interrupt signal is raised from
\texttt{MandelbrotWidget}. The accelerated version can be easily derived
by applying the methodology in
Sec.~\ref{sec:selfoffloading}. Figure~\ref{fig:mandelbrot} presents
experimental results obtained by running the original code and the \ff
accelerated version for 2, 4, 8 and 16 threads. As shown in the figure, the
application has been tested for 8 refinement passes of the pixmap (according to the
original algorithm) in 4 different regions of the plane exhibiting
different execution times (and different regularity); in terms of Amdahl's law, the smaller
this time, the smaller the fraction of the application that can be made parallel, the smaller the maximum speedup.  As is clear from the figure, the \ff
accelerator is able to boost the sequential application close to
ideal speedup in almost all cases.

\subsection{N-queens problem}

The N-queens problem is a generalization of the well-known 8-queens
problem. N-queens have to be placed on an NxN sized chessboard such
that no queen can attack any of the others. The objective is to count all
possible solutions.
One of the fastest sequential implementations available for solving the
problem is the heavily optimised C code written by Jeff Somers \cite{somers}.
Somer's algorithm calculates one half of the solutions (considering one half of the
columns), then flips the results over the ``Y axis'' of the board.
Every solution can be reflected that way to generate another unique solution.
That is because a solution cannot be symmetrical across the Y axis.

We attempted to accelerate the execution time of the sequential code using FastFlow.
The FastFlow version uses the farm construct without the collector
entity. A stream of independent tasks, each corresponding to an initial
placement of a number of queens on the board, is produced and offloaded
into the farm accelerator.
The placement of the remaining queens in a task is handled by one of the
accelerator's worker threads.
In order to speed up the code, we simply applied the methodology described
in Sec. \ref{sec:selfoffloading}. We copied the part of code that we
wished to accelerate into the \emph{svc} method of the Worker class;
defined the stream type in such a way that it contained all the local variables
that must be passed to the worker thread for the computation; and produced the
stream of tasks from the initial placement of a given number of queens.
No additional data structure or optimization has been added to the new code
version.

\begin{table}[tb]
\label{fig:nqueentable}
\begin{center}
\textsf{\scriptsize
\begin{tabular}{rrrrrrr}
\toprule
\multicolumn{6}{c}{Andromeda: 8-core 16-hyperthreads platform}\\
\midrule
Board size & \# of solutions & Seq. Time & \ff Time & \# of tasks & Speedup\\
\cmidrule{2-6}
18x18&       666,090,624&      5:53&	       34 & 1710 & 10.4\\
19x19&	   4,968,057,848&     44:56&	     4:23 & 2072 & 10.2\\
20x20&	  39,029,188,884&   6:07:21&        35:41 & 2482 & 10.3\\
21x21&	 314,666,222,712& $\sim$ 2.2days& 5:07:19 & 2943 & 10.3\\
\bottomrule
\end{tabular}\\[2ex]
\begin{tabular}{rrrrrrr}
\toprule
\multicolumn{6}{c}{Ottavinareale: 8-core platform}\\
\midrule
Board size & \# of solutions & Seq. Time & \ff Time & \# of tasks & Speedup\\
\cmidrule{2-6}
18x18&       666,090,624&      6:52&	   1:06 & 1710 & 6.24\\
19x19&	   4,968,057,848&     53:28&	   8:26 & 2072 & 6.34\\
20x20&	  39,029,188,884&   7:16:27&     1:8:56 & 2482 & 6.52\\
21x21&	 314,666,222,712& $\sim$ 2.7days&  9:48:28 & 2943  & 6.69\\
\bottomrule
\end{tabular}
}
\end{center}
\caption{N-queens execution time breakdown on two different multi-core
  platforms
  (Andromeda and Ottavinareale).\label{tab:nqueens}}
\end{table}

Table \ref{tab:nqueens}  shows the execution times for the original sequential  and the FastFlow
accelerated versions for different board sizes. In all the tests we used 16 worker threads and
the stream has been produced from the initial placement of 4 queens (the resulting
number of tasks is shown in the table).
As can be seen, more than 10x speedup in the execution time has been obtained
without any particular code optimization.

\section{Related Work}
\label{sec:relwork}

In computing the word accelerator is used to refer to mechanisms that
are used to speed up computation.
The most widespread accelerators are
hardware ones: the standard CPU is coupled with dedicated hardware
optimized for a specific kind of computation. Examples include
cryptographic accelerators,  which have been developed to perform
processor-intensive decryption/encryption;   TCP/IP Offload Engines,
which process the entire TCP/IP stack; and finally the well-known
Graphics Processing Units (GPUs), which initially targeted graphics
computations and are now increasingly used for a wider range of
computationally intensive applications. Usually accelerators feature a
different architecture with respect to standard CPUs and thus, in
order to ease exploitation of their computational power, specific
libraries are developed. In the case of GPUs those libraries include
\emph{Brook}, NVidia \emph{CUDA} and \emph{OpenCL}.

Brook \cite{Brook} provides extensions to the C language with single
program multiple data (SPMD) operations on streams. It abstracts the
stream hardware as a coprocessor to the host system.
User defined functions operating on stream elements are called \emph{kernels} and
can be executed in parallel.
Brook kernels feature blocking behaviour: the execution of a kernel
must complete before the next kernel can execute.
A similar execution model is available on
GPUs via the OpenCL framework \cite{opencl} and CUDA \cite{CUDA}. \ff
accelerator   differs from that of the previous libraries
because it does not target specific accelerators; instead it make
possible the usage of some of the cores as a virtual accelerator.

A  recent work \cite{charm++Acc}, using the Charm++ programming
model \cite{Productivity02}, has demonstrated that accelerator
extensions are able to obtain good performance. Furthermore, code
written with these extensions is portable without changing the
application's source code. However, in order to exploit the
accelerator features, the application has to be entirely rewritten
using the Charm++ framework; this is not necessary in \ff.

Stream processing is extensively discussed in literature. Stream
languages are often
motivated by the application style used in image processing,
networking, media processing, and a wide and growing number of
problems in finance.

StreamIt \cite{streamIt} is an explicitly parallel programming
language based on the Synchronous Data Flow model.  A program is
represented as a set of filters, i.e. autonomous actors (containing
Java-like code) that communicate through first-in first-out (FIFO)
data channels.  Filters can be assembled in \emph{pipeline}, possibly
with a \emph{FeedbackLoop}, or according to a \emph{SplitJoin}
data-parallel schema.

S-Net \cite{snet} is a coordination language to describe the
communications of asynchronous sequential components (a.k.a. boxes)
written in a sequential language (e.g. C, C++, Java) through typed
streams.  The overall design of S-Net is geared towards facilitating
the composition of components developed in isolation.

Streaming applications are also targeted by TBB
\cite{intel:skeletons:tbb} through the \emph{pipeline}
construct. However, TBB does not support any kind of non-linear
streaming network, which therefore has to be embedded in a pipeline
with significant drawbacks in terms of expressivity and
performance. As an example, a streaming network structured a workflow
(a direct acyclic graph, actually) can be embedded in pipeline but
this require pipeline stages to bypass data in which they have no
interest. This clearly requires to change both the interfaces of the
stages and their business logic and can be hardly made parametric.
In addition, artificial data dependencies are (uselessly) introduced
in the application with the consequent performance drawback.

\emph{OpenMP} \cite{openMP} is a very popular thread-based framework
for multi-core architectures. It mostly targets Data Parallel
programming and provides means to easily incorporate threads into
sequential applications at a relatively high level.  In an OpenMP
program data needs to be labeled as shared or private, and compiler
directives have to be used to annotate the code.

Both OpenMP and TBB can be used to accelerate serial C/C++ programs in
specific portions of code, even if they do not natively include farm
skeletons, which are instead realised by using lower-level features
such as the \emph{task} annotation in OpenMP and the
\emph{parallel\_for} construct in TBB.  OpenMP does not require
restructuring of the sequential program, while with TBB, which
provides thread-safe containers and some parallel algorithms, it is
not always possible to accelerate the program without some refactoring
of the sequential code.

In our vision, FastFlow falls between the easy programming of OpenMP
and the powerful mechanisms provided by TBB. The FastFlow accelerator
allows one to speed-up execution of a wide class of existing C/C++
serial programs with just minor modifications to the code. To the best
of our knowledge none of the mentioned frameworks supports lock-free
(and CAS-free) synchronizations.

\section{Conclusions}
In this paper we introduced  the \ff accelerator which represents a further
evolution of the \ff framework specifically designed to support the
semi-automatic parallelization of existing  sequential C/C++
applications on multi-cores. The \ff accelerator exhibits well-defined
functional and extra-functional behaviour represented by a skeleton
composition; this helps in ensuring the correctness of the
parallelization process. The main vehicle of parallelization is offloading of
code kernels onto a number of additional threads on the same CPU; we
call this  technique \emph{self-offloading}.

All in all, the work addresses an increasingly crucial problem for
modern software engineering: how to make existing applications capable of
effectively using modern multi-core systems with limited human effort. In
this the \ff accelerator is supported by a semi-formal methodology and by
the unique ability of \ff to support very fine grain tasks on standard
multi-cores.

The effectiveness of the proposed methodology has been demonstrated by simple
but challenging applications. The \ff library and the code for all the
applications in Sec.~\ref{sec:exp} are available under GPL at the \ff
website~\cite{fastflow:web}.


\begin{thebibliography}{10}

\bibitem{lithium:sem:CLSS}
M.~Aldinucci and M.~Danelutto.
\newblock Skeleton based parallel programming: functional and parallel semantic
  in a single shot.
\newblock {\em Computer Languages, Systems and Structures}, 33(3-4):179--192,
  Oct. 2007.

\bibitem{beske:ipdps:09}
M.~Aldinucci, M.~Danelutto, and P.~Kilpatrick.
\newblock Autonomic management of non-functional concerns in distributed and
  parallel application programming.
\newblock In {\em Proc. of Intl. Parallel \& Distributed Processing Symposium
  (IPDPS)}, pages 1--12, Rome, Italy, May 2009. IEEE.

\bibitem{fastflow:parco:09}
M.~Aldinucci, M.~Danelutto, M.~Meneghin, P.~Kilpatrick, and M.~Torquati.
\newblock Efficient streaming applications on multi-core with {FastFlow}: the
  biosequence alignment test-bed.
\newblock In {\em Proc. of Parallel Computing (ParCo)}, Lyon, France, Sept.
  2009.

\bibitem{fastflow:pdp:10}
M.~Aldinucci, M.~Meneghin, and M.~Torquati.
\newblock Efficient {Smith-Waterman} on multi-core with {FastFlow}.
\newblock In {\em Proc. of Intl. Euromicro PDP 2010: Parallel Distributed and
  network-based Processing}, Pisa, Italy, Feb. 2010. IEEE.

\bibitem{fastflow:web}
M.~Aldinucci and M.~Torquati.
\newblock {\em FastFlow website}, 2010.
\newblock \url{http://mc-fastflow.sourceforge.net/}.

\bibitem{patterson:cacm:09}
K.~Asanovic, R.~Bodik, J.~Demmel, T.~Keaveny, K.~Keutzer, J.~Kubiatowicz,
  N.~Morgan, D.~Patterson, K.~Sen, J.~Wawrzynek, D.~Wessel, and K.~Yelick.
\newblock A view of the parallel computing landscape.
\newblock {\em CACM}, 52(10):56--67, 2009.

\bibitem{Brook}
I.~Buck, T.~Foley, D.~Horn, J.~Sugerman, K.~Fatahalian, M.~Houston, and
  P.~Hanrahan.
\newblock Brook for {GPUs}: stream computing on graphics hardware.
\newblock In {\em ACM SIGGRAPH '04 Papers}, pages 777--786, New York, NY, USA,
  2004. ACM Press.

\bibitem{cole:manifesto:02}
M.~Cole.
\newblock Bringing skeletons out of the closet: A pragmatic manifesto for
  skeletal parallel programming.
\newblock {\em Parallel Computing}, 30(3):389--406, 2004.

\bibitem{fastforward:ppopp:08}
J.~Giacomoni, T.~Moseley, and M.~Vachharajani.
\newblock Fastforward for efficient pipeline parallelism: a cache-optimized
  concurrent lock-free queue.
\newblock In {\em Proc. of the 13th ACM SIGPLAN Symposium on Principles and
  practice of parallel programming (PPoPP)}, pages 43--52, New York, NY, USA,
  2008. ACM.

\bibitem{ibm:offload:09}
IBM Corp.
\newblock {\em IBM Dynamic Application Virtualization}, 2010.
\newblock \url{http://www.alphaworks.ibm.com/tech/dav}.

\bibitem{intel:skeletons:tbb}
Intel Corp.
\newblock {\em Threading Building Blocks}, 2009.
\newblock \url{http://www.threadingbuildingblocks.org/}.

\bibitem{Productivity02}
L.~V. Kal{\'e}.
\newblock Performance and productivity in parallel programming via processor
  virtualization.
\newblock In {\em Proc. of the 1st Intl. Workshop on Productivity and
  Performance in High-End Computing (at HPCA 10)}, Madrid, Spain, Feb. 2004.

\bibitem{opencl}
Khronos Compute Working Group.
\newblock {\em OpenCL}, Nov. 2009.
\newblock \url{http://www.khronos.org/opencl/}.

\bibitem{CUDA}
D.~Kirk.
\newblock {NVIDIA} {CUDA} software and {GPU} parallel computing architecture.
\newblock In {\em Proc. of the 6th Intl. Symposium on Memory Management
  (ISMM)}, pages 103--104, New York, NY, USA, 2007. ACM.

\bibitem{charm++Acc}
D.~M. Kunzman and L.~V. Kal\'{e}.
\newblock Towards a framework for abstracting accelerators in parallel
  applications: experience with cell.
\newblock In {\em SC '09: Proc. of the Conference on High Performance Computing
  Networking, Storage and Analysis}, pages 1--12, New York, NY, USA, 2009. ACM.

\bibitem{lamport:bakery}
L.~Lamport.
\newblock A new solution of dijkstra's concurrent programming problem.
\newblock {\em Commun. ACM}, 17(8):453--455, 1974.

\bibitem{Lamport}
L.~Lamport.
\newblock Specifying concurrent program modules.
\newblock {\em ACM Trans. Program. Lang. Syst.}, 5(2):190--222, 1983.

\bibitem{ABA:98}
M.~M. Michael and M.~L. Scott.
\newblock Nonblocking algorithms and preemption-safe locking on multiprogrammed
  shared memory multiprocessors.
\newblock {\em Journal of Parallel and Distributed Computing}, 51(1):1--26,
  1998.

\bibitem{qt:web}
Nokia Corp.
\newblock {\em Qt – Cross-platform application and {UI} framework}, 2010.
\newblock \url{http://qt.nokia.com/}.

\bibitem{openMP}
I.~Park, M.~J. Voss, S.~W. Kim, and R.~Eigenmann.
\newblock Parallel programming environment for {OpenMP}.
\newblock {\em Scientific Programming}, 9:143--161, 2001.

\bibitem{blog:acm:reed:2009}
D.~Reed.
\newblock {\em High-Performance Computing: Where'd The Abstractions Go?}
\newblock BLOG@CACM, May 2009.

\bibitem{snet}
A.~Shafarenko, C.~Grelck, and S.-B. Scholz.
\newblock Semantics and type theory of {S-Net}.
\newblock In {\em Proc. of the 18th Intl. Symposium on Implementation and
  Application of Functional Languages (IFL'06)}, TR 2006-S01, pages 146--166.
  E\"otv\"os Lor\'and University, Faculty of Informatics, Budapest, Hungary,
  2006.

\bibitem{somers}
J.~Somers.
\newblock {\em The N Queens Problem: a study in optimization}, 2010.
\newblock \url{http://jsomers.com/nqueen_demo/nqueens.html}.

\bibitem{streamIt}
W.~Thies, M.~Karczmarek, and S.~P. Amarasinghe.
\newblock {StreamIt}: A language for streaming applications.
\newblock In {\em Proc. of the 11th Intl. Conference on Compiler Construction
  (CC)}, pages 179--196, London, UK, 2002. Springer-Verlag.

\bibitem{ske:wikipedia}
Wikipedia.
\newblock {\em Algorithmic skeleton}, 2009.
\newblock \url{http://en.wikipedia.org/wiki/Algorithmic_skeleton}.

\end{thebibliography}

\end{document}